\documentclass[nature,floatfix,twocolumn,showpacs,preprintnumbers,superscriptaddress,amsmath,amssymb]{revtex4-1}

\usepackage{graphicx}  
\usepackage{dcolumn}   
\usepackage{bm}        
\usepackage{xcolor}
\usepackage[normalem]{ulem}

\begin{document}

\title[]{Magnetic on-off switching of a plasmonic laser}

\author{Francisco Freire-Fern\'{a}ndez}
\email{francisco.freire.fernandez@northwestern.edu}
\affiliation{Department of Applied Physics, Aalto University School of Science, P.O. Box 15100, FI-00076 Aalto, Finland}
\affiliation{Department of Chemistry, Northwestern University, Evanston, Illinois 60208, United States}	
\author{Javier Cuerda}
\affiliation{Department of Applied Physics, Aalto University School of Science, P.O. Box 15100, FI-00076 Aalto, Finland}
\author{Konstantinos S. Daskalakis}
\affiliation{Department of Applied Physics, Aalto University School of Science, P.O. Box 15100, FI-00076 Aalto, Finland}
\affiliation{Department of Mechanical and Materials Engineering, Faculty of Technology, University of Turku, 20014 Turun Yliopisto, Finland}
\author{Sreekanth Perumbilavil}
\affiliation{Department of Applied Physics, Aalto University School of Science, P.O. Box 15100, FI-00076 Aalto, Finland}
\author{Jani-Petri Martikainen}
\affiliation{Department of Applied Physics, Aalto University School of Science, P.O. Box 15100, FI-00076 Aalto, Finland}
\author{Kristian Arjas}
\affiliation{Department of Applied Physics, Aalto University School of Science, P.O. Box 15100, FI-00076 Aalto, Finland}
\author{P\"{a}ivi T\"{o}rm\"{a}}
\email{paivi.torma@aalto.fi}
\affiliation{Department of Applied Physics, Aalto University School of Science, P.O. Box 15100, FI-00076 Aalto, Finland}	\author{Sebastiaan van Dijken}
\email{sebastiaan.van.dijken@aalto.fi}
\affiliation{Department of Applied Physics, Aalto University School of Science, P.O. Box 15100, FI-00076 Aalto, Finland}
	
\date{\today}	

\maketitle	

\textbf{The nanoscale mode volumes of surface plasmon polaritons have enabled plasmonic lasers and condensates with ultrafast operation~\cite{Oulton2009,Hill2014,wang_structural_2018,hakala_bose-einstein_2018}. Most plasmonic lasers are based on noble metals, rendering the optical mode structure inert to external fields. Here, we demonstrate active magnetic-field control over lasing in a periodic array of Co/Pt multilayer nanodots immersed in an IR-140 dye solution.  We exploit the magnetic nature of the nanoparticles combined with mode tailoring to control the lasing action. Under circularly polarized excitation, angle-resolved photoluminescence measurements reveal a transition between lasing action and non-lasing emission as the nanodot magnetization is reversed. Our results introduce magnetization as a means of externally controlling plasmonic nanolasers, complementary to the modulation by excitation~\cite{Knudson2019}, gain medium~\cite{Yang2015c,Taskinen2020a} or substrate~\cite{Wang2018a}. Further, the results show how effects of magnetization on light that are inherently weak can be observed in the lasing regime, inspiring studies of topological photonics~\cite{Haldane2008,bahari_nonreciprocal_2017,ozawa_topological_2018}.}

Plasmonic nanostructures feature electromagnetic modes that confine light into sub-wavelength volumes. Consequently, the Purcell effect is strong, and when such structures are paired with emitters, enhancement of both spontaneous and stimulated emission can be achieved. Plasmonic lasers have been realized with architectures ranging from single nanoparticles and metal-insulator thin films to random, aperiodic, periodic and superperiodic arrays of nanoparticles~\cite{Oulton2009,Hill2014,wang_structural_2018,Zhou2013,Schokker2016,Wang2017,ramezani_plasmon-exciton-polariton_2017}. While the effects of the nanoparticle shape and arrangement have been studied extensively~\cite{Knudson2019,Wang2018a,Wang2017}, the material choice has been limited mostly to noble metals, except for some recently proposed alternatives~\cite{Ha2018,Pourjamal2019}. The selection of noble metals, and in particular dielectric alternatives, is motivated by the minimization of ohmic losses typical for plasmonic materials. Lower damping leads to a higher mode quality factor (Q-factor) and a stronger Purcell effect, which benefits lasing. Due to inherently high ohmic losses, nanostructures made of magnetic materials have been largely overlooked as a platform for plasmonic lasers, even when they would offer, in principle, the powerful possibility of modifying the optical modes by a magnetic field during device operation~\cite{Pourjamal2019}. Here, we demonstrate that this unique advantage can be experimentally realized.  
   
\begin{figure*}
\includegraphics{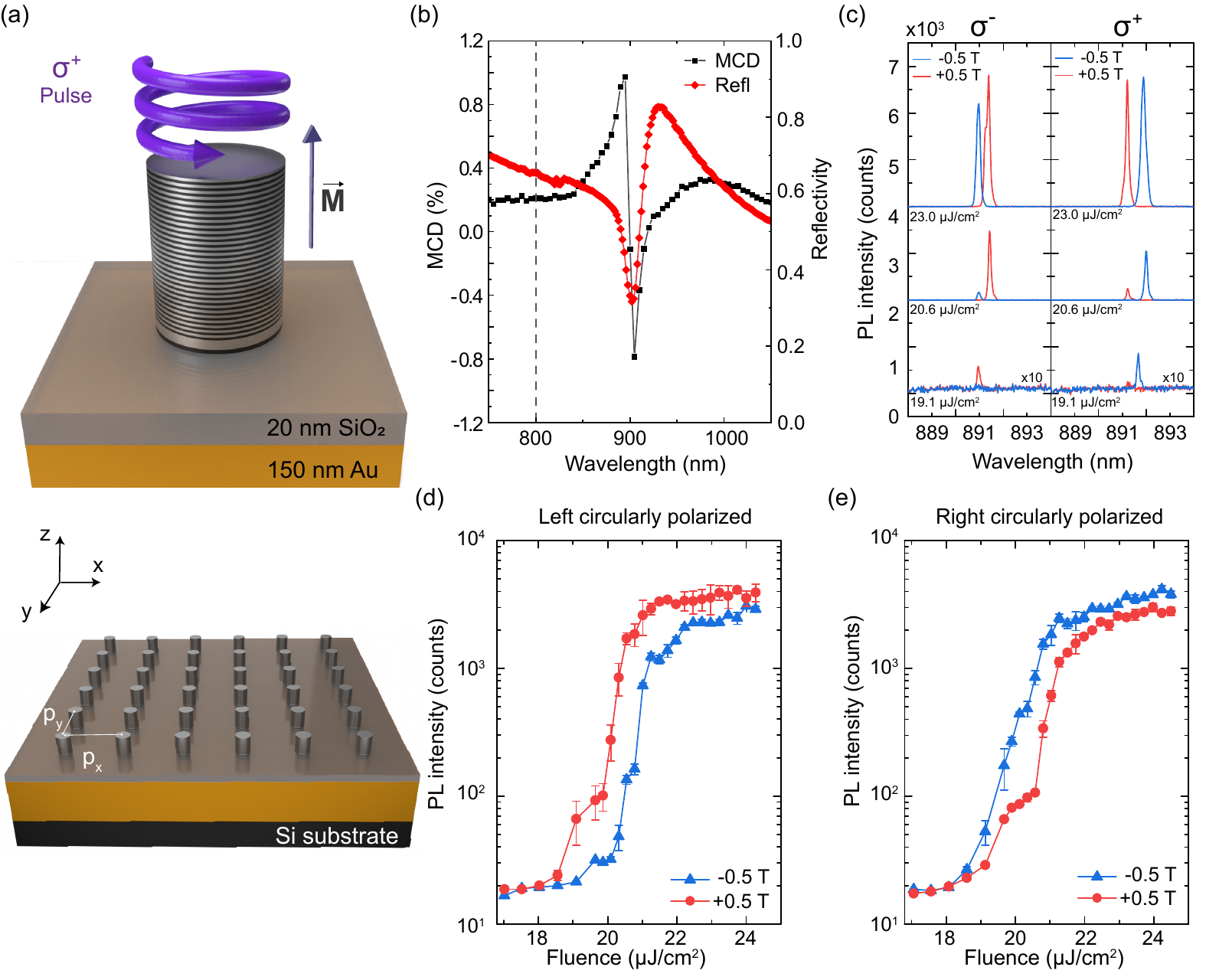}
\caption{\label{fig:1}Magnetic-field control of plasmonic lasing in a square array of Co/Pt nanodots. (a) Co/Pt nanodots (upper panel) with diameter $D$ = 220 nm, a height of 68 nm, and perpendicular magnetization are arranged in periodic lattices ($p_{x}$ = $p_{y}$ = 590 nm) on top of a Au/Si$\textnormal{O}_{2}$ bilayer (lower panel). (b) Experimental MCD and reflectivity spectra. Both spectra show prominent features at the SLR wavelength determined by the lattice period. The MCD spectrum is obtained by subtracting the reflectivity spectra of $\sigma^{+}$ light for up and down magnetization and signal normalization. The dashed line marks the wavelength of the pump pulses in the lasing experiments. (c) PL intensity spectra recorded normal to the sample plane using $\sigma^{-}$ and $\sigma^{+}$ excitation for three pump fluences. For clarity, the PL intensity spectra corresponding to 19.1 $\mu\textnormal{J/cm}^{2}$ have been multiplied by a factor of 10 and each fluence pair is offset vertically by 2000 counts. (d),(e) PL intensity as a function of the pump laser fluence for (d) $\sigma^{-}$ and (e) $\sigma^{+}$ excitation. In (c)-(e), the blue and red data are recorded while a $-$0.5 T and +0.5 T magnetic field saturates the perpendicular magnetization of the Co/Pt nanodots down and up, respectively.}
\end{figure*}
	
Contrary to noble metals, the permittivity tensor of magnetic metals contains non-zero off-diagonal components that depend on the direction of magnetization. Consequently, magnetic switching alters the optical response of magnetic materials, giving rise to magneto-optical phenomena such as the Faraday effect, magneto-optical Kerr effect, and magnetic circular dichroism (MCD). While these effects have been exploited successfully in the field of plasmonics for biosensing applications utilizing phase-sensitive detection~\cite{Maccaferri2015}, the absolute modulation of the light intensity through magneto-optics is weak. For instance, the polarization of light reflected from or transmitted through arrays of magnetic nanodots rotates by up to one degree~\cite{FreireFernandez2020}, which corresponds to an intensity modulation of the order 1\%. Here we show that, although small, magneto-optical effects lead to subtle changes in the mode structure of the array, and these, in turn, become prominently visible due to  non-linearities inherent in a lasing process, enabling full on-off switching of the lasing action by an external magnetic field.  

To demonstrate magnetic-field control of plasmonic lasing, we fabricated square and rectangular arrays of Co/Pt multilayer nanodots on a Au/Si$\textnormal{O}_{2}$ bilayer and immersed the structures in a gain medium consisting of a 12 mM IR-140 dye solution (see Methods). Figure \ref{fig:1}(a) shows the full  $\textnormal{Ta}(2)/\textnormal{Pt}(4)/[\textnormal{Co}(1)/\textnormal{Pt}(1)]_{30}/\textnormal{Pt}(2)$ nanodot multilayer stack, where the $[\textnormal{Co}(1)/\textnormal{Pt}(1)]$ bilayer repeats 30 times and the numbers in parentheses indicate the layer thickness in nanometers. Hereafter, we refer to these structures as Co/Pt nanodots. Co/Pt multilayers were selected as the nanodot material because of abrupt magnetic switching in a perpendicular magnetic field and full remanence (Supplementary Fig.~1), enabling non-volatile magnetic-field control of plasmonic lasing. Nanodots of diameter $D$ = 220 nm were arranged in periodic lattices as shown in Fig. \ref{fig:1}(a) with periods $p_{x}=p_{y}$ = 590 nm for square arrays and $p_{y}$ ranging from 520 nm to 540 nm in 5 nm steps for rectangular arrays. Arranging metallic nanostructures into periodic lattices enables the excitation of surface lattice resonances (SLRs), which are hybridized modes of the lattice diffracted orders and localized surface plasmons (LSPs) of the nanodots~\cite{Zou2004a}. The experimental reflectivity spectrum of the square array ($p_{x}=p_{y}$= 590 nm) displays a pronounced minimum at the SLR wavelength (Fig.~\ref{fig:1}(b)), in agreement with finite-element method (FEM) simulations (Supplementary Fig.~2). The SLR of the nanodot array overlaps with the emission spectrum of the gain medium (Supplementary Fig.~3), which is essential for optical feedback. Moreover, the quality factor of the SLR mode compares to that of noble-metal systems~\cite{FreireFernandez2020}, providing high rates of stimulated emission.

Lasing is achieved when the square Co/Pt nanodot array is excited by linearly, left circularly ($\sigma^{-}$), or right circularly ($\sigma^{+}$) polarized 200 fs pulses at 800 nm, where the chirality is defined from the receiver's point of view (Fig. \ref{fig:1}(a)). As an example, we show photoluminescence (PL) intensity spectra for $\sigma^{-}$ and $\sigma^{+}$ pulses in Fig. \ref{fig:1}(c). The strong non-linear increase and narrowing of the emission peak signifies lasing above a critical pump fluence. The intensity of the lasing peak increases by about two orders of magnitude and its full width at half maximum (FWHM) is $\sim$0.2 nm (Supplementary Fig.~4). More importantly, for circularly polarized pump pulses, the lasing threshold and lasing intensity depend on the direction of magnetization in the Co/Pt nanodots. For $\sigma^{-}$ excitation, the threshold fluence is smallest when the magnetization points up (+0.5 T) and largest if the magnetization points down ($-$0.5 T). Consequently, by selecting a pump fluence just above the lower threshold, we are able to drastically alter the lasing intensity through magnetic switching (75 - 90\% intensity modulation at 20.6 $\mu\textnormal{J/cm}^{2}$ in Fig. \ref{fig:1}(c)). 
We note that the lasing threshold in our system is one order of magnitude lower than that of other plasmonic array- and dye-based lasers, being more similar to the values reported for high-gain materials such as quantum dots~\cite{Guan2020}. The low threshold fluence could be due to an enhanced dye emission rate caused by the proximity of the molecules to the laser-irradiated nanodots and the Au film~\cite{Lawley1980}. 

Figure \ref{fig:1}(d) summarizes the variation of the PL intensity with pump fluence and the direction of magnetization for $\sigma^{-}$ pulses. The difference in threshold fluence for up and down magnetization is $\sim$0.8 $\mu\textnormal{J/cm}^{2}$ (4\%). The magnetization state also affects the lasing intensity in saturation. Similar results are obtained under $\sigma^{+}$ excitation, but with a reversed dependence on the direction of perpendicular magnetization (Fig. \ref{fig:1}(e)). Additionally, switching the magnetization state of the Co/Pt nanodots does not change the lasing intensity when using linearly polarized pulses (Supplementary Fig.~5).

\begin{figure}
\includegraphics{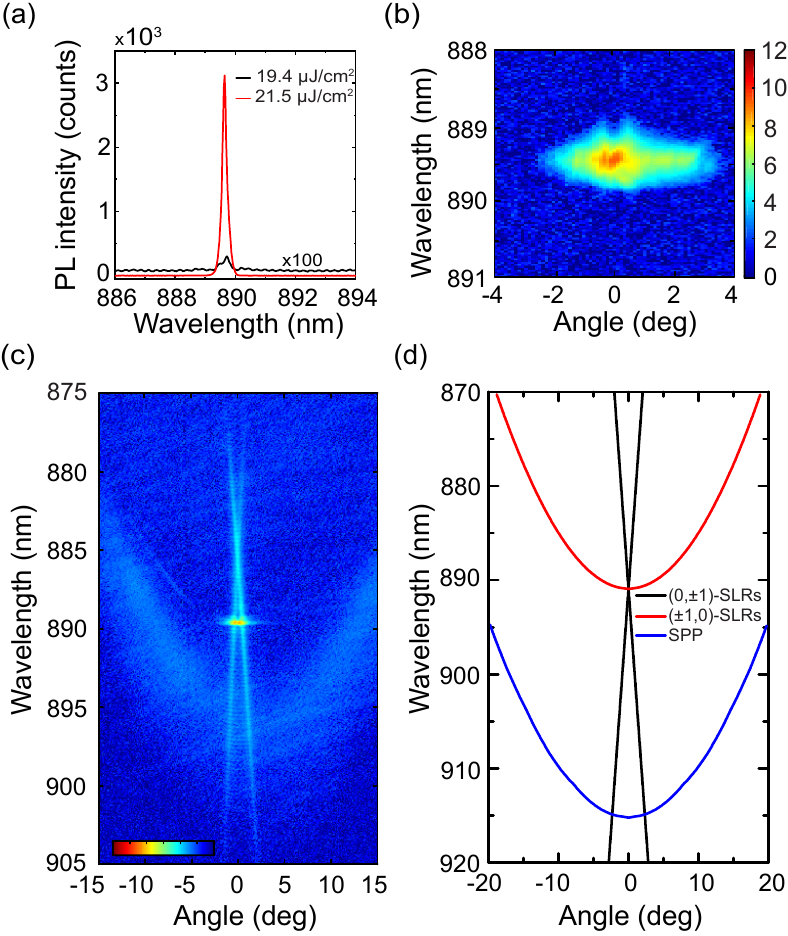}
\caption{\label{fig:2} Analysis of the lasing mode for a square Co/Pt nanodot array. (a) PL intensity spectra recorded normal to the sample plane for a pump fluence close to and above the lasing threshold. (b),(c) Angle-resolved emission spectra recorded at (b) 21.5 $\mu\textnormal{J/cm}^{2}$ and (c) 19.4 $\mu\textnormal{J/cm}^{2}$. The color bars are identical and use a logarithmic scale. (d) Calculated dispersion bands for a sample with a square nanodot array. All the data are obtained for normal-incident light with linear polarization along the $x$-axis. The angle corresponds to lattice momentum along the $y$-direction.} 
\end{figure}

Next, we discuss angle-resolved PL measurements to identify the lasing mode. We consider data recorded close to and above the lasing threshold, as indicated in Fig. \ref{fig:2}(a). The above-threshold data shown in Fig. \ref{fig:2}(b) demonstrates highly directional lasing normal to the sample plane at 889.3 nm. The angle-resolved emission spectrum obtained near the lasing threshold (Fig. \ref{fig:2}(c)) reveals the band dispersion of the plasmonic system. Calculations based on the empty-lattice approximation shown in Fig. \ref{fig:2}(d) reproduce the main modes: SLRs supported by the square Co/Pt nanodot array and a SPP excited at the Au/Si$\textnormal{O}_{2}$ interface via grating coupling. For normal-incident light with linear polarization along the $x$-axis, the modes with linear dispersion correspond to SLRs excited through the (0,$\pm$1) diffracted orders whereas the red parabolic band belongs to SLRs excited through the ($\pm$1,0) diffracted orders (see Methods). The broad parabolic band in Fig. \ref{fig:2}(c) and the blue curve in Fig.~\ref{fig:2}(d) depict the SPP mode. From the analysis in Fig. \ref{fig:2}, we conclude that stimulated emission to modes that are located close to the diffracted orders at the $\Gamma$-point (lattice momenta $k_x=k_y=0$) produces lasing in our system. A more accurate determination of the lasing modes including their chirality is given below, after we consider the effect of magnetic field on optical absorption. 

Circularly polarized light excites the free electrons of a metal nanodot into spectrally degenerate rotational motion. When a magnetic field is applied or the nanodot exhibits a net magnetization, the Lorentz force on the electrons lifts the degeneracy~\cite{Pineider2013}. Consequently, the optical response of a magnetic nanodot resides at slightly different frequencies for $\sigma^-$ and $\sigma^+$ excitation and the two frequencies interchange when an external magnetic field switches the magnetization. A similar MCD effect also appears in the SLR modes of a magnetic nanodot lattice~\cite{Kataja2016b}, as illustrated for the square array in Fig. \ref{fig:1}(b). The change in optical absorption due to the MCD effect could increase the excitation of dye molecules in the vicinity of the nanodot lattice. This effective increase in pumping would shift the lasing threshold curve, but only by the amount of increased absorption, i.e. linearly. The MCD effect at the pump wavelength is below 1\%, whereas the threshold fluence of the square array changes by about 4\% (Figs. \ref{fig:1}(d),(e)). Besides, a mere shift of the threshold curve does not produce a higher saturated PL intensity, in contradiction to our observation. Non-linear changes of the saturated PL intensity and the lasing threshold can, instead, be caused by changes in the lasing mode lifetime, as well as in the so-called $\beta$-factor, which quantifies which portion of the total emission of the gain medium goes to the lasing mode. To estimate such non-linear effects, we now analyze in detail the properties of the lasing modes, as well as other modes overlapping with the emission spectrum of the molecules.

\begin{figure*}
\includegraphics[width=0.99\textwidth]{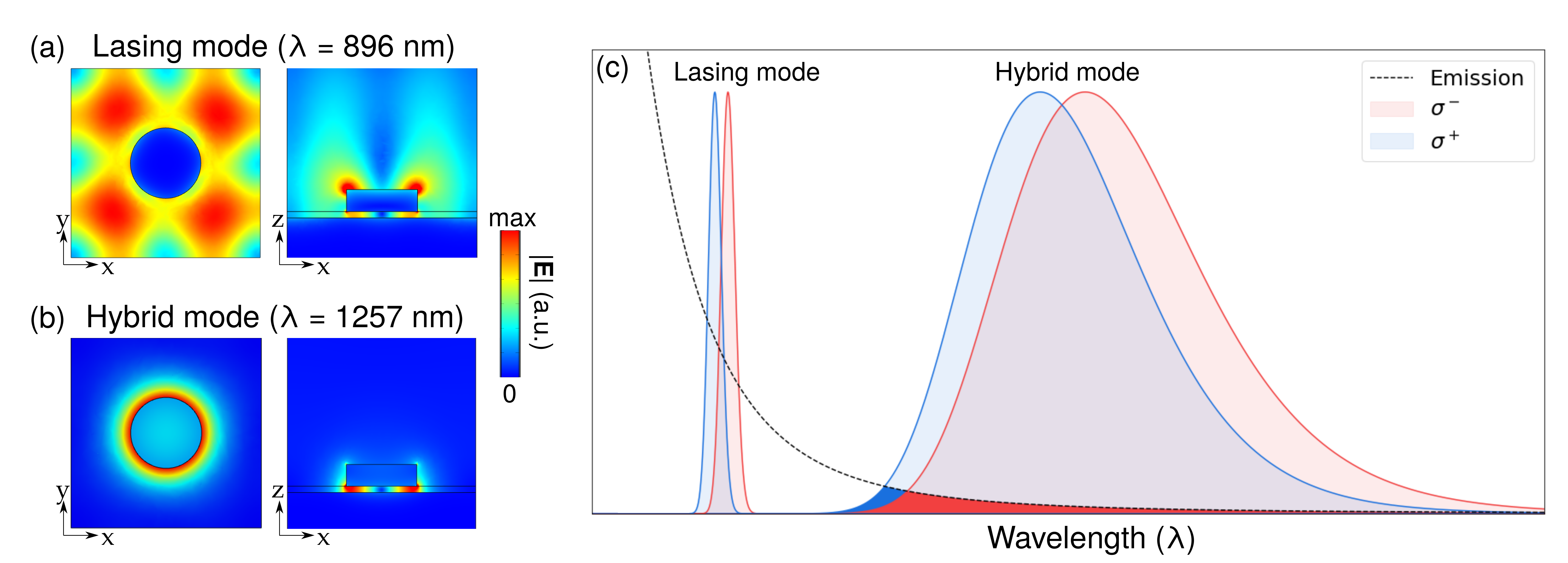}
\caption{\label{fig:3}Chiral modes emerging in a square lattice of Co/Pt nanodots. (a) FEM electric field profile of the out-of-plane SLR lasing mode. In the presence of perpendicular magnetization, the degenrate SLR doublet splits into $\sigma^+$ and $\sigma^-$ modes (see (c)). (b) FEM field profile of the hybrid mode (in-plane SLR with strong LSP contribution and hybridization to SPPs). The degeneracy of this mode is also lifted by the perpendicular magnetization. The modes in (a) and (b) are $\sigma^+$-polarized, but the $\sigma^-$ modes look similar. (c) Illustration of the overlap between the emission spectrum of the dye solution (dashed line) and the lasing and hybrid modes. The hybrid mode at lower wavelength has a bigger overlap with the emission spectrum of the dye and, consequently, the gain available for the lasing mode of the same chirality is lower. The schematic is not in scale but it depicts the mode ordering as obtained from the simulations for magnetization pointing up.    
} 
\end{figure*}

The lasing mode identified in Fig. \ref{fig:2} turns out to be an out-of-plane SLR mode related to the ($\pm$1,0) and (0,$\pm$1) diffracted orders (Supplementary Fig.~6). In a square lattice, this mode is doubly degenerate due to the $x$-$y$ symmetry. FEM simulations of nanodot arrays with perpendicular magnetization show that the degeneracy is lifted, producing two modes of $\sigma^-$ and $\sigma^+$ character at wavelengths 896.5 nm and 895.8 nm, respectively. The $\sim$5 nm difference to the experimental lasing peaks is understandable as the refractive index of the dye solution may slightly change upon pumping (note that the $n=1.48$ used in the simulations produces an excellent match with the experimental reflectivity spectra taken without pumping (Supplementary Fig.~2)). The two modes and their field profile are shown in Fig.~\ref{fig:3}(a),(c) (more details in Supplementary Fig.~7). In nanoparticle array lasers, the pump often excites not only the gain medium but also slightly the plasmonic modes, and this small excitation then stimulates emission, causing lasing polarization to be dependent on pump polarization~\cite{wang_structural_2018}. It can thus be expected that $\sigma^-$ and $\sigma^+$ pump pulses trigger lasing in the corresponding modes. Indeed, this is what we observe in the experiments shown in Fig.~\ref{fig:1}(c) and Supplementary Fig.~8: there are two lasing peaks, and the peaks interchange when either the magnetization direction or pump helicity switches as expected for modes of opposite chirality. We find good agreement with the magnitude of the splitting between the two modes given by simulations, namely 0.7 nm, and the experimentally observed distance between the lasing peaks, 0.46 $\pm$ 0.25 nm (average over 6 measurements). Further evidence of the chiral doublet is given by the fact that, consistent with circularly polarized emission, we observe both $x$- and $y$-polarized components in the lasing emission (Supplementary Fig.~9). The observation of two chiral lasing modes is a remarkable result, since such splitting of degenerate modes by time-reversal symmetry breaking (magnetization direction) is needed in creating topologically non-trivial systems~\cite{Haldane2008}. Here we show that effects of magnetic time-reversal symmetry breaking in plasmonics, which are small in general due to the weakness of magneto-optical effects, become prominent through suitable mode design and visible in the lasing regime.   

\begin{figure*}
\includegraphics{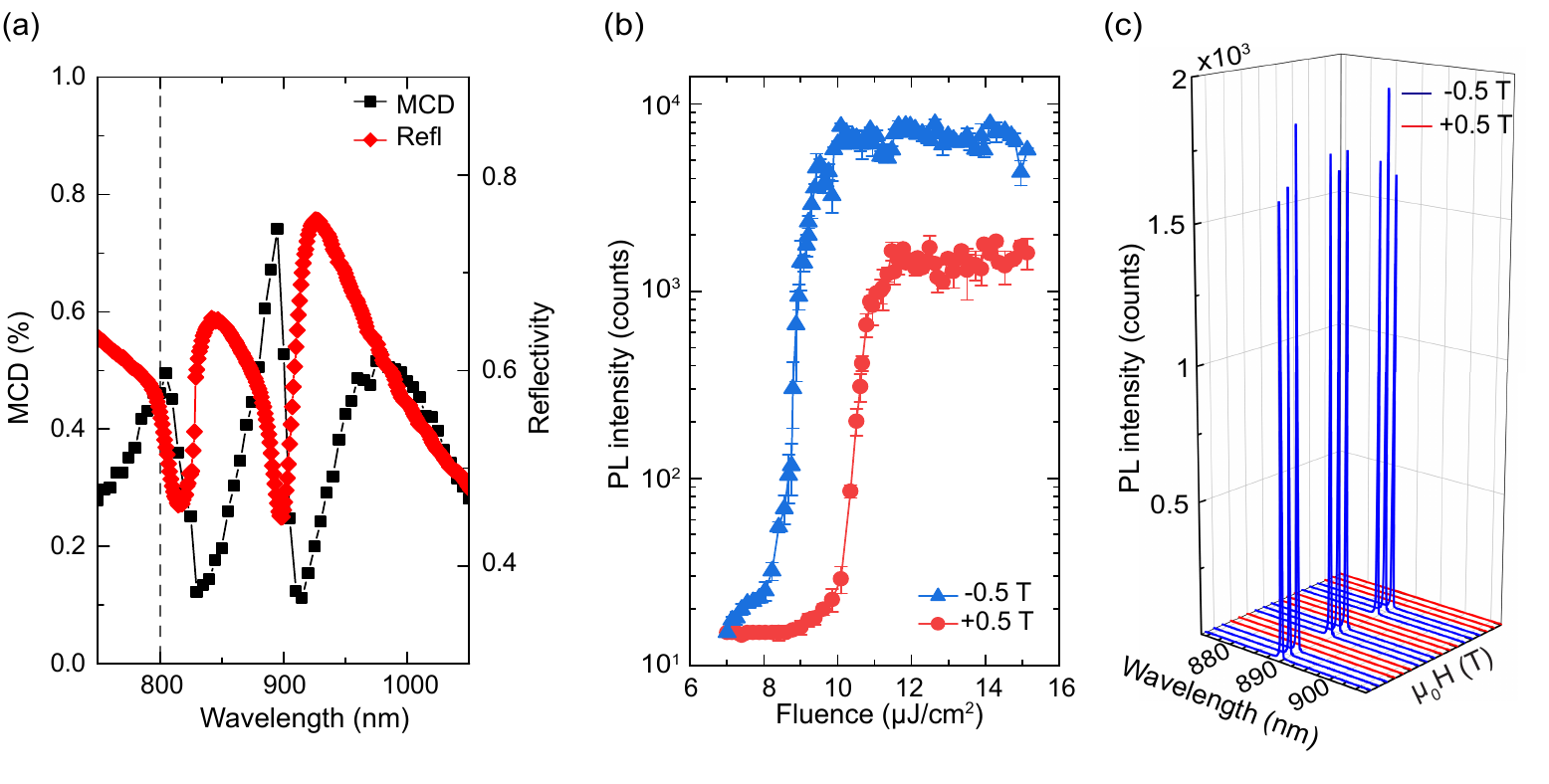}
\caption{\label{fig:4} Magnetic-field control of plasmonic lasing in rectangular arrays of Co/Pt nanodots. (a) MCD and reflectivity spectra of a rectangular array with $p_{x}$ = 590 nm and $p_{y}$ = 530 nm for $\sigma^{+}$ excitation. (b) PL intensity as a function of the pump laser fluence for an array with $p_{y}$ = 530 nm and $\sigma^{+}$ excitation. Results recorded with the magnetization of the Co/Pt nanodots pointing up and down are shown in red and blue. (c) PL intensity recorded at a $\sigma^{+}$ pump fluence of 9.8 $\mu\textnormal{J/cm}^{2}$ while repeatedly switching the magnetization of the Co/Pt nanodots between down (blue) and up (red).}
\end{figure*}

According to the FEM simulations, the $\sigma^-$ and $\sigma^+$ lasing modes have lifetimes too similar to explain the notable modulation of the lasing threshold. One should bear in mind, however, that the pump may provide a small seed excitation not only to the lasing mode but also to other modes. Stimulated emission to those modes can then deplete the gain available for the lasing mode. Using FEM simulations, we have identified a broad hybrid mode at higher wavelength corresponding to an in-plane SLR with a strong LSP component. The SPPs of the gold film underneath are involved also, as the mode does not appear at this wavelength when the gold is absent (in that case, the in-plane SLR is close to the diffracted orders). Figure \ref{fig:3}(b) shows how the field of the hybrid mode is strongly enhanced close to the nanodots. Just like the lasing mode, this mode is doubly degenerate in the non-magnetized case, and splits into $\sigma^-$ and $\sigma^+$ modes at wavelengths 1270 nm and 1257 nm, respectively, in the presence of perpendicular magnetization (Supplementary Figs.~6 and 7). Although these broad modes are spectrally far from the lasing wavelength, they still overlap with the emission spectrum of the molecules. Consequently, the pump can excite these modes in a chirality-dependent manner, which triggers stimulated emission either to the $\sigma^-$ or $\sigma^+$ mode. This emission reduces the gain available for the lasing mode, which can be described as an effective, chirality-dependent reduction of the lasing mode $\beta$-factor. Based on the wavelengths and linewidths of the modes obtained from FEM simulations and the observed emission spectrum (Supplementary Fig.~7), we estimate a 1 - 2\% change in the lasing mode $\beta$-factor, which in turn changes the threshold and PL intensity non-linearly. For realistic parameters, 3 - 4\% changes in the lasing threshold are simulated for this gain competition effect (Supplementary Figs.~10 and 11), which is comparable to our experimental observations. Direct experimental support for this scenario is given by Fig.~\ref{fig:1}(c), always showing a lower threshold and higher PL for the lasing peak at higher wavelength. This is consistent with the spectral ordering of the simulated modes depicted in Fig.~\ref{fig:3}(c): the hybrid mode which is farther away from the lasing modes, and thus reduces the gain less, has the same chirality as the higher-wavelength lasing mode, explaining its lower threshold and higher PL. We note that $\sigma^-$/$\sigma^+$ helicity are defined the same way in the experiments and simulations, as confirmed by the similar behavior of the simulated MCD (Supplementary Fig. 12) and the measured data (Fig.~\ref{fig:1}(b)). 

The proposed switching mechanism relies on stimulated emission, as spontaneous emission does not prefer the chirality of one mode above the other. To further confirm the mechanism and to exploit it in designing even stronger magnetic field effects, one could maximize the amount of pump energy that is coupled directly to the plasmonic modes. This can be achieved by fabricating rectangular arrays where one of the periods is selected to generate a SLR mode at the excitation wavelength (800 nm) and the other, as for the square array, produces a SLR within the emission band of the dye molecules ($\sim$890 nm). We accomplished this by varying $p_{y}$ from 520 nm to 540 nm in 5 nm steps while keeping $p_{x}$ fixed at 590 nm. Figure \ref{fig:4}(a) depicts the MCD and reflectivity spectra of the rectangular array with $p_{y}=530$ nm. 
Compared to the square array (Fig. \ref{fig:1}(b)), the rectangular array absorbs about 60\% more light at the pump wavelength (see reflectivity data in Fig.~\ref{fig:4}(a)). As demonstrated by Fig.~\ref{fig:4}(b), the lasing threshold approximately halves. Further, because the difference in absorption of $\sigma^-$ and $\sigma^+$ light is larger in the rectangular array (the MCD effect is enhanced by about 50\% compared to the square array), the chirality-dependent coupling to the broad hybrid modes changes accordingly. The lattice-designed increase in MCD enhances the change in lasing threshold when the magnetization is switched.
Moreover, the hybrid modes in the rectangular array are pulled closer to the lasing mode according to our simulations (wavelengths 1266 nm for $\sigma^-$ and 1253 nm for $\sigma^+$ light, for $p_{x}=590$ nm and $p_{y}=530$ nm, Supplementary Figs.~7 and 13), which further increases the chirality-dependent gain depletion effect. Indeed, the experimentally observed changes in the lasing action are larger than in the case of the square array (Fig.~\ref{fig:4}(b)). We estimate a 2.5 - 4\% change in the effective $\beta$-factor, which leads to 6-26\% change in the threshold for typical $\beta$-factors (Supplementary Fig.~14). Data for other rectangular arrays demonstrating a direct correlation between the optical absorption of $\sigma^-$ and $\sigma^+$ light at the excitation wavelength and the magnitude of the magnetic switching effect are shown in Supplementary Fig.~15. In the rectangular lattice, the energy difference of the SLR modes related to the (±1,0) and (0,±1) diffracted orders breaks the degeneracy of the lasing mode. Consequently, a chiral doublet does not form in the presence of magnetization. FEM simulations indeed show that the lasing mode is $y$-polarized (Supplementary Fig. 13). This is consistent with the $y$-polarized lasing emission observed in our experiments (Supplementary Fig. 9), in clear contrast to the square lattice case.  Lasing results for all rectangular arrays are summarized in Supplementary Figs.~16-20.

Tailoring of the SLR modes at the excitation wavelength and at a wavelength that overlaps with the emission spectrum of the gain medium provides full on-off switching of the lasing signal by an external magnetic field, as demonstrated in Fig.~\ref{fig:4}(c). Here, the modulation of the PL intensity exceeds two orders of magnitude at a pump fluence of 9.8 $\mu\textnormal{J/cm}^{2}$. While our proof-of-concept experiments are conducted by placing permanent magnets near the plasmonic sample, magnetic switching is inherently fast, a feature that is critical for non-volatile magnetic data storage technology~\cite{Vedmedenko_2020}. Moreover, ultrafast all-optical helicity-dependent magnetic switching has been demonstrated for ferromagnetic metals, including Co/Pt~\cite{Lambert2014}. The modulation of lasing intensity through magnetic switching, as demonstrated here, therefore provides a feasible pathway towards actively controlled coherent light sources for enhanced light-matter interactions on the nanoscale. Our results hold promise also for topological photonics~\cite{ozawa_topological_2018,Ota2020}. Photonic analogues of topological insulators can, in theory, be realized with time reversal symmetry breaking provided by material magnetization~\cite{Haldane2008}. However, the effect is weak at visible frequencies and external magnetic fields are needed for topological lasing~\cite{bahari_nonreciprocal_2017}. We have discovered that the interplay of magnetized plasmonic nanodots with the symmetry of the array can lead to a splitting of the chiral modes by about 0.5 nm or more, which is remarkable compared e.g.~to the 42 pm topological bandgap in Ref.~\cite{bahari_nonreciprocal_2017}. The splits are below the natural linewidth of the modes but become visible in the lasing regime as we have shown. Our results suggest magnetic nanodot arrays as an exciting platform for studies of topological photonics; with permanently magnetized nanodots, a topological system could be realized even without an external magnetic field.   

\section*{Methods}

\subsection*{Sample fabrication}
The plasmonic structures were fabricated on Si substrates with a native Si$\textnormal{O}_{2}$ layer. First, a 2 nm Ti/150 nm Au film was grown by electron beam evaporation. The metal film was then covered by 20 nm Si$\textnormal{O}_{2}$ using atomic layer deposition.
On top of the Si$\textnormal{O}_{2}$ layer, Co/Pt nanodot arrays were patterned by electron beam lithography in a Vistec EBPG5000pES system. After defining holes in a PMMA resist layer, a 2 nm Ta/4 nm Pt/[1 nm Co/1 nm Pt]$_{30}$/2 nm Pt multilayer stack was grown by magnetron sputtering. The resist layer was lift-off in acetone. The array size was chosen to be 0.5 mm by 0.5 mm as this is approximately the minimum size required to measure their magneto-optical response in our home-built magneto-optical setup. Nonetheless, lasing from smaller arrays is expected~\cite{Wang2020c}. 

\subsection*{Optical and magneto-optical characterization}
The optical and MCD response of the plasmonic structures was characterized in reflection using a supercontinuum laser (NKT SuperK EXW-12) and a photodector (Hinds DET-200-002). The samples were placed between the poles of an electromagnet (GMW 3470) for the application of perpendicular magnetic fields. All optical measurements were conducted near normal incidence through a hole in one of the pole pieces. Reflectivity spectra were recorded in zero magnetic field using linearly polarized light. A quarter-wave plate was used to circularly polarize ($\sigma^{-}$ or $\sigma^{+}$) the incident laser beam in the MCD measurements. MCD spectra were obtained by normalizing the difference in optical reflectivity for up and down magnetization to the reflectivity measured in zero external field. The magnetization of the Co/Pt nanodots was saturated up or down by a $\pm$0.5 T magnetic field. The Co/Pt nanodots were immersed in IR-140 dye solution using a cover glass, to ensure an identical dielectric environment as in the lasing experiments.

\subsection*{Angle-resolved PL measurements}
Angle-resolved PL measurements were performed using a 12 mM IR-140 dye solution. The IR-140 dye (Sigma-Aldrich) was dissolved in a 1:2 solution of dimethyl sulfoxide and benzyl alcohol to match the refractive index of the cover glass. The plasmonic structures were excited at 800 nm and normal incidence by 200 fs pulses from a Ti:sapphire laser (Coherent). The repetition rate was 1 kHz. We focused the laser beam to a diameter of 650 $\mu$m to fully irradiate the $500\times500$ $\mu$m$^2$ nanodot arrays. The PL of the arrays was collected using a 10$\times$ CFI plan fluor objective (Nikon) with a numerical aperture of 0.3. The objective back focal plane was imaged onto the entrance slit of a spectrometer (SP2500 Princeton Instruments) and projected onto a 2D CCD detector (PIXIS 400, Princeton instruments), providing wavelength and angle-resolved images (Fig. \ref{fig:2}(b),(c)). From these images, we extracted the single-pixel maximum intensity counts within the spectral range where lasing was observed to study the dependence of PL intensity on pump fluence. Three consecutive images were recorded and the extracted maximum PL intensity was averaged. The spectral resolution of the CCD camera was 0.14 nm. Lasing experiments were conducted for linearly, left circularly ($\sigma^{-}$), and right circularly ($\sigma^{+}$) polarized laser pulses. The magnetization of the Co/Pt nanodots was switched by rotating a permanent magnet behind the sample. The field at the sample location was $\pm$0.5 T. 

\subsection*{Empty-lattice model}
We used an analytical model based on the empty-lattice approximation~\cite{Kichin2012,Guo2017} to identify the modes present in the angle-resolved emission spectrum in Fig.~\ref{fig:2}(c). In this approach, the plasmon modes in the nanodot lattice are defined as~\cite{Guo2017,Cherqui2019}:
\begin{equation}
	|\mathbf{k}_{||}+\mathbf{G}|=\sqrt{\epsilon}\frac{\omega}{c},
	\label{eq:2}
\end{equation}
where $\mathbf{k}_{||}$ is the in-plane wave vector, $\mathbf{G}=n_{x}G_{x}\hat{\mathbf{x}}+n_{y}G_{y}\hat{\mathbf{y}}$, $G_{x,y}=2\pi/p_{x,y}$ are the reciprocal vectors of the lattice, $n_{x,y}$ are integers (0,$\pm$1,...) corresponding to the diffracted orders, $\epsilon$ is the dielectric constant of the surrounding environment, $c$ is the speed of light, and $\omega$ is the angular frequency. In our experiments, the spectrometer entrance slit was parallel to the $y$-axis and $\mathbf{k}_{||}=\mathbf{k}_{y}$, with $\mathbf{k}_{x}=0$. Hence,
\begin{equation}
	|\mathbf{k}_{y}+\mathbf{G}|=\sqrt{\epsilon}\frac{\omega}{c}
	\label{eq:3}
\end{equation}
By definition, the $\Gamma-$point is the frequency at which the diffracted orders $n_{x,y}=\pm 1$ intersect for $\mathbf{k}_{||}=0$ in Eq.~(\ref{eq:2}). Therefore, a four-fold degeneracy arises at that frequency.  According to Eq.~(\ref{eq:3}), the frequencies of the SLRs corresponding to the diffracted orders (0,$\pm 1$) depend linearly on the in-plane wave vector, resulting in two linear dispersion bands (black curves in Fig. \ref{fig:2}(d)). Conversely, the SLRs corresponding to the diffracted orders ($\pm 1,0$) yield two degenerated parabolic dispersion bands (red curve in Fig. \ref{fig:2}(d)). 

For the evanescent SPP modes propagating at the 150 nm Au/20 nm Si$\textnormal{O}_{2}$ interface, the empty-lattice model yields the relation~\cite{Kichin2012}:

\begin{equation}
	|\mathbf{k}_{||}+\mathbf{G}|=k_{SPP}=\frac{\omega}{c}\sqrt{\frac{\epsilon_{m}\epsilon_{d}}{\epsilon_{m}+\epsilon_{d}},}
	\label{eq:4}
\end{equation}
where $k_{SPP}$ is the momentum of the SPP mode and $\epsilon_m$ and $\epsilon_{d}$ are the dielectric constant of Au and Si$\textnormal{O}_{2}$, respectively. Equation (\ref{eq:4}) also results in a parabolic variation of the frequency with in-plane wave vector. Consequently, the SPPs exhibit parabolic dispersion bands (blue curve in Fig.~\ref{fig:2}(d)) that are slightly redshifted compared to the parabolic SLR bands. Because of the transversal-magnetic nature of the SPP modes, the counterpart of the linear SLR dispersion bands is not present in this case. A refractive index $n=1.51$ was used here to make the empty lattice model match with the experiments shown in Fig.~\ref{fig:2}; the actual SLR modes are typically slightly away from the empty lattice approximation.

\subsection*{Rate equation analysis}
In order to model the interplay of the emission of the gain medium with the plasmonic modes supported by the magnetic nanodot lattice, we implement a standard rate equation approach, as in Ref.~\cite{daskalakis_ultrafast_2018}. We describe the molecules as optically pumped four-level quantum emitters, with time-dependent populations from $N_{0}$ to $N_{3}$ in increasing order of energy, given by:
\begin{align}
\dfrac{dN_{3}}{dt}=&r(t)(N_{0}-N_3)-\dfrac{N_{3}}{\tau_{32}}\label{rateq1}\\
\dfrac{dN_{2}}{dt}=&-\beta n_{ph}\dfrac{\left(N_{2}-N_{1}\right)}{\tau_{21}}-\dfrac{N_{2}}{\tau_{21}}-\dfrac{N_{2}}{\tau_{20}}+\dfrac{N_{3}}{\tau_{32}}\label{rateq2}\\
\dfrac{dN_{1}}{dt}=&\beta n_{ph}\dfrac{\left(N_{2}-N_{1}\right)}{\tau_{21}}+\dfrac{N_{2}}{\tau_{21}}-\dfrac{N_{1}}{\tau_{10}}\label{rateq3}\\
\dfrac{dN_{0}}{dt}=&-r(t)(N_{0}-N_3)+\dfrac{N_{2}}{\tau_{20}}+\dfrac{N_{1}}{\tau_{10}}\label{rateq4}
\end{align}
where $r$ is the time-dependent pumping rate corresponding to a pulsed pump excitation, with: $r(t)=rf(t)$ and $f(t)=\exp(-(t-t_0)^2/\tau_p^2)$, so that $t_0\gg \tau_p$ and $\tau_p=200$ fs (same as the pump pulse length in the experiment); the lifetimes $\tau_{32}=500$ fs, $\tau_{20}=240$ ps, and $\tau_{10}=4$ ps account for non-radiative transitions, and $\tau_{21}=720$ ps is the lifetime of the lasing transition. For more details and definitions see the supplemental material of Ref.~\cite{daskalakis_ultrafast_2018}. The photon number $n_{ph}$ generated by both spontaneous and stimulated emission is determined by the expression:
\begin{equation}\label{rateq5}
\dfrac{dn_{ph}}{dt}=\beta n_{ph}\dfrac{\left(N_{2}-N_{1}\right)}{\tau_{21}}+\beta\dfrac{N_{2}}{\tau_{21}}-\dfrac{n_{ph}}{\tau_{cav}}  ,
\end{equation}
where $\tau_{cav}$ is the cavity lifetime, and the $\beta$-factor accounts for the fraction of emitted photons that decay into the lasing mode. We simulated the PL intensity and threshold curve by solving the time evolution of the coupled equations (\ref{rateq1})-(\ref{rateq5}), using different values of the pump intensity $r$, and integrating $n_{ph}$ over a long time. The lifetimes of the lasing modes ($\tau_{cav}$) in our magnetic system were calculated from numerical FEM simulations of the electromagnetic fields. Further, we used the overlap of the molecule emission spectrum with the lasing and hybrid plasmonic modes (Fig.~\ref{fig:3}(c)) as an input for estimating the $\beta$-factor. More details on these calculations are given in the Supplementary Information.

\subsection*{Estimation of the $\beta$-factors}

We consider gain competition between the lasing and hybrid modes, and mark the decay of the emission to these modes by $\gamma_L$ (the lasing mode) and $\gamma_{-,hybrid}$, $\gamma_{+,hybrid}$ (the hybrid modes); $\gamma_i$ denotes any of these, with $i$ specifying the mode. We estimate the rates $\gamma_i$ as
\begin{equation}
    \gamma_i = \int_0^{E_{cutoff}} p(E) \alpha^2 \sigma_i(E) dE,
\end{equation}
where $E_{cutoff}$ is the energy below which the lasing and hybrid modes are located (Supplementary Fig.~7), $p(E)$ is the normalized emission spectrum, $\alpha^2$ is the projection of mode to $\sigma^-$/$\sigma^+$ basis (0.5 for the linear polarization lasing mode of the rectangular case, 1 for other modes) and $\sigma_i(E)$ is the absorption probability of mode $i$. The emission spectrum is approximated with a Lorentzian fitted to the measured emission spectrum (Supplementary Figs.~3 and 7). The absorption spectra $\sigma_i(E)$ are estimated by Lorentzians with parameters (peak energy and linewidth) obtained from the FEM simulations (Supplementary Fig.~7).

The $\beta$-factor is defined as the proportional decay rate into the selected mode~\cite{des2016plasmonic}, where the total decay includes other decay channels:
\begin{equation}
    \beta_i 
   = \frac{\gamma_i}{\gamma_{tot}}  = \frac{\gamma_i}{\gamma_i + \gamma_{other}}.
\end{equation}
If $\gamma_i$, $\gamma_j$ $<< \gamma_{other}$, then $\beta_i/\beta_j \simeq \gamma_i/\gamma_j$. This is the case here as the $\beta$-factors of the lasing and hybrid modes are very small (e.g., $\sim10^{-3}$).

We assume that the lasing and hybrid modes compete about the same gain, and their $\beta$-factors are of the same order of magnitude. Then if the $\beta$-factor of the hybrid mode changes, there is a corresponding change in the lasing mode $\beta$-factor. That is, decreased/increased amount of emission to the hybrid mode is given to/taken away from the lasing mode. We estimate this change $\Delta$ in $\frac{\beta_-}{\beta_+} = 1 + \Delta$ by
\begin{equation}
   \Delta = \frac{\gamma_{-,hybrid}}{\gamma_L} \left( 1 - \frac{\gamma_{-,hybrid}}{\gamma_{+,hybrid}} \right).
\end{equation}
This corresponds to a change in the $\beta$ of the hybrid mode, scaled by how significant the hybrid mode decay is compared to the lasing mode ($\frac{\gamma_{-,hybrid}}{\gamma_L}$, which is of order one).

\subsection*{Finite element analysis}
In order to theoretically calculate the effect of nanodot magnetization, and, in particular, to provide the wavelength and lifetime of the plasmonic modes in the system for the rate equations, we carried out numerical simulations of the electromagnetic fields using the finite element method (FEM). We modelled a three-dimensional realistic structure consisting of an infinite lattice of Co/Pt nanodots combined with the Au layer (as in Fig.~\ref{fig:1}(a)) by simulating one unit cell and imposing periodic boundary conditions in the $x-$ and $y-$directions. We used the geometrical parameters from the experiment, and described the local optical response of the Co/Pt material through a non-diagonal permittivity tensor \cite{Zvezdin1997}:
\begin{align}\label{matrix4modes}
\epsilon=
\begin{pmatrix}
\epsilon_{xx} & \epsilon_{xy} & 0 \\
\epsilon_{yx} & \epsilon_{yy} & 0 \\
0 & 0 & \epsilon_{zz} ,
\end{pmatrix}
\end{align}   
where $\epsilon_{xx}=\epsilon_{yy}=\epsilon_{zz}=\epsilon_{d}$, $\epsilon_{xy}=-\epsilon_{yx}=i\epsilon_{0}$, and $\epsilon_{d}$, $\epsilon_{o}$ are complex-valued parameters coming from previously measured data \cite{Sato_1992}. Further, we model the metal response of the Au film using a complex frequency-dependent isotropic permittivity available in experimental tabulated data \cite{JohnsonChristy} and assume that both the spacer that separates the metallic substrate from the nanodots, and the background medium, have a refractive index of $n=1.48$.

The FEM analysis was performed by combining eigenmode studies, and frequency-domain scattering simulations. Eigenmode simulations allow us to identify the modes supported by the system without any external excitation, including dark modes, and provide their near fields together with a complex eigenfrequency: $f=f'+if''$. The energy, linewidth and lifetime of the mode are then given by $E=hf'$, $\Gamma=hf''$, and $\tau=1/(2\pi f'')$, where $h$ is Planck's constant. The results of these simulations are shown in Fig.~3(a),(b), and Supplementary Figs.~6 and 13. In Figs.~\ref{fig:3}(a),(b) we show results for one unit cell at planes intersecting the nanodot at its mid-height (left panel) and across its diameter (right panel).

In addition, we calculated the optical reflectivity and MCD spectra of the system using frequency-domain scattering simulations with normal-incident plane wave illumination. Simulated optical reflectivity and MCD spectra are shown in Supplementary Figs.~2 and 12.

\subsection*{Identifying the chirality of the eigenmodes}

Finite element eigenmode simulations provide the energy, lifetime, and electromagnetic fields of all the magnetized and non-magnetized modes in the system. However, identifying the polarization of the resulting modes is not straightforward as these simulations do not consider the explicit coupling to external illumination. We develop a theoretical method to unveil the chirality of the in-plane and out-of-plane modes involved in our analysis.

Considering that the magnetized modes emerge from a doublet of $x-$ and $y-$polarized modes in the square array (see Supplementary Fig.~6(b)-(e)), we define our magnetized states according to the field distributions of the linearly polarized modes:
\begin{equation}
|m_{1}\rangle=C_{x,m_{1}}|\leftrightarrow\rangle+C_{y,m_{1}}|\updownarrow\rangle
\end{equation}
\begin{equation}
|m_{2}\rangle=C_{x,m_{2}}|\leftrightarrow\rangle+C_{y,m_{2}}|\updownarrow\rangle
\end{equation}
with 
\begin{equation}
C_{x,m_{1,2}}=\langle\leftrightarrow|m_{1,2}\rangle
\end{equation}
\begin{equation}
C_{y,m_{1,2}}=\langle\updownarrow|m_{1,2}\rangle ,
\end{equation}
where the scalar product is defined as the overlap integral
\begin{equation}
\langle i|j\rangle=\int_{\textrm{Unit Cell}}d^{3}\mathbf{r}\ \mathbf{u}_{i}^{\ast}(\mathbf{r})\cdot\mathbf{u}_{j}(\mathbf{r})
\end{equation}
with the normalization
\begin{equation}
\mathbf{u}_{i}(\mathbf{r})=\dfrac{\mathbf{E}_{i}(\mathbf{r})}{\sqrt{\int_{\textrm{Unit Cell}}d^{3}\mathbf{r}|\mathbf{E}_{i}(\mathbf{r})|^{2}}}   .
\end{equation}
Moreover, we can project the $x, y-$linearly polarized basis onto the $\sigma^+$ and $\sigma^-$ polarized basis, obtaining 
\begin{equation}
|m_{1}\rangle=C_{\sigma^{-},m_{1}}|\sigma^{-}\rangle+C_{\sigma^{+},m_{1}}|\sigma^{+}\rangle
\end{equation}
\begin{equation}
|m_{2}\rangle=C_{\sigma^{-},m_{2}}|\sigma^{-}\rangle+C_{\sigma^{+},m_{2}}|\sigma^{+}\rangle
\end{equation}
with
\begin{equation}
C_{\sigma^{-},m_{1,2}}=\dfrac{1}{\sqrt{2}}(C_{x,m_{1,2}}-iC_{y,m_{1,2}})
\end{equation}
\begin{equation}
C_{\sigma^{+},m_{1,2}}=\dfrac{1}{\sqrt{2}}(C_{x,m_{1,2}}+iC_{y,m_{1,2}}).
\end{equation}
These amplitudes, and their modulus, reveal whether the mode is $\sigma^+$ or $\sigma^-$ polarized. \\

\section*{Data availability}
The data from this work can be obtained from the corresponding authors upon reasonable request.

\bibliography{Lasing_paper_2,bec_paper2019,Refs_BGV_3,Refs}

\begin{thebibliography}{10}
\expandafter\ifx\csname url\endcsname\relax
  \def\url#1{\texttt{#1}}\fi
\expandafter\ifx\csname urlprefix\endcsname\relax\def\urlprefix{URL }\fi
\providecommand{\bibinfo}[2]{#2}
\providecommand{\eprint}[2][]{\url{#2}}

\bibitem{Oulton2009}
\bibinfo{author}{Oulton, R.~F.} \emph{et~al.}
\newblock \bibinfo{title}{{Plasmon lasers at deep subwavelength scale}}.
\newblock \emph{\bibinfo{journal}{Nature}} \textbf{\bibinfo{volume}{461}},
  \bibinfo{pages}{629--632} (\bibinfo{year}{2009}).

\bibitem{Hill2014}
\bibinfo{author}{Hill, M.~T.} \& \bibinfo{author}{Gather, M.~C.}
\newblock \bibinfo{title}{{Advances in small lasers}}.
\newblock \emph{\bibinfo{journal}{Nature Photon.}}
  \textbf{\bibinfo{volume}{8}}, \bibinfo{pages}{908--918}
  (\bibinfo{year}{2014}).

\bibitem{wang_structural_2018}
\bibinfo{author}{Wang, D.}, \bibinfo{author}{Wang, W.},
  \bibinfo{author}{Knudson, M.~P.}, \bibinfo{author}{Schatz, G.~C.} \&
  \bibinfo{author}{Odom, T.~W.}
\newblock \bibinfo{title}{Structural engineering in plasmon nanolasers}.
\newblock \emph{\bibinfo{journal}{Chem. Rev.}} \textbf{\bibinfo{volume}{118}},
  \bibinfo{pages}{2865--2881} (\bibinfo{year}{2018}).

\bibitem{hakala_bose-einstein_2018}
\bibinfo{author}{Hakala, T.~K.} \emph{et~al.}
\newblock \bibinfo{title}{{Bose-{{Einstein}} condensation in a plasmonic
  lattice}}.
\newblock \emph{\bibinfo{journal}{Nature Phys.}} \textbf{\bibinfo{volume}{14}},
  \bibinfo{pages}{739} (\bibinfo{year}{2018}).

\bibitem{Knudson2019}
\bibinfo{author}{Knudson, M.~P.} \emph{et~al.}
\newblock \bibinfo{title}{{Polarization-dependent lasing behavior from
  low-symmetry nanocavity arrays}}.
\newblock \emph{\bibinfo{journal}{ACS Nano}} \textbf{\bibinfo{volume}{13}},
  \bibinfo{pages}{7435--7441} (\bibinfo{year}{2019}).

\bibitem{Yang2015c}
\bibinfo{author}{Yang, A.} \emph{et~al.}
\newblock \bibinfo{title}{{Real-time tunable lasing from plasmonic nanocavity
  arrays}}.
\newblock \emph{\bibinfo{journal}{Nat. Commun.}} \textbf{\bibinfo{volume}{6}},
  \bibinfo{pages}{6939} (\bibinfo{year}{2015}).

\bibitem{Taskinen2020a}
\bibinfo{author}{Taskinen, J.~M.} \emph{et~al.}
\newblock \bibinfo{title}{{All-optical emission control and lasing in plasmonic
  lattices}}.
\newblock \emph{\bibinfo{journal}{ACS Photonics}} \textbf{\bibinfo{volume}{7}},
  \bibinfo{pages}{2850--2858} (\bibinfo{year}{2020}).

\bibitem{Wang2018a}
\bibinfo{author}{Wang, D.} \emph{et~al.}
\newblock \bibinfo{title}{{Stretchable nanolasing from hybrid quadrupole
  plasmons}}.
\newblock \emph{\bibinfo{journal}{Nano Lett.}} \textbf{\bibinfo{volume}{18}},
  \bibinfo{pages}{4549--4555} (\bibinfo{year}{2018}).

\bibitem{Haldane2008}
\bibinfo{author}{Haldane, F.~D.} \& \bibinfo{author}{Raghu, S.}
\newblock \bibinfo{title}{{Possible realization of directional optical
  waveguides in photonic crystals with broken time-reversal symmetry}}.
\newblock \emph{\bibinfo{journal}{Phys. Rev. Lett.}}
  \textbf{\bibinfo{volume}{100}}, \bibinfo{pages}{013904}
  (\bibinfo{year}{2008}).

\bibitem{bahari_nonreciprocal_2017}
\bibinfo{author}{Bahari, B.} \emph{et~al.}
\newblock \bibinfo{title}{Nonreciprocal lasing in topological cavities of
  arbitrary geometries}.
\newblock \emph{\bibinfo{journal}{Science}} \textbf{\bibinfo{volume}{358}},
  \bibinfo{pages}{636--640} (\bibinfo{year}{2017}).

\bibitem{ozawa_topological_2018}
\bibinfo{author}{Ozawa, T.} \emph{et~al.}
\newblock \bibinfo{title}{{Topological photonics}}.
\newblock \emph{\bibinfo{journal}{Rev. Mod. Phys.}}
  \textbf{\bibinfo{volume}{91}}, \bibinfo{pages}{015006}
  (\bibinfo{year}{2019}).

\bibitem{Zhou2013}
\bibinfo{author}{Zhou, W.} \emph{et~al.}
\newblock \bibinfo{title}{{Lasing action in strongly coupled plasmonic
  nanocavity arrays}}.
\newblock \emph{\bibinfo{journal}{Nat. Nanotechnol.}}
  \textbf{\bibinfo{volume}{8}}, \bibinfo{pages}{506--511}
  (\bibinfo{year}{2013}).

\bibitem{Schokker2016}
\bibinfo{author}{Schokker, A.~H.} \& \bibinfo{author}{Koenderink, A.~F.}
\newblock \bibinfo{title}{{Lasing in quasi-periodic and aperiodic plasmon
  lattices}}.
\newblock \emph{\bibinfo{journal}{Optica}} \textbf{\bibinfo{volume}{3}},
  \bibinfo{pages}{686} (\bibinfo{year}{2016}).

\bibitem{Wang2017}
\bibinfo{author}{Wang, D.} \emph{et~al.}
\newblock \bibinfo{title}{{Band-edge engineering for controlled multi-modal
  nanolasing in plasmonic superlattices}}.
\newblock \emph{\bibinfo{journal}{Nat. Nanotechnol.}}
  \textbf{\bibinfo{volume}{12}}, \bibinfo{pages}{889--894}
  (\bibinfo{year}{2017}).

\bibitem{ramezani_plasmon-exciton-polariton_2017}
\bibinfo{author}{Ramezani, M.} \emph{et~al.}
\newblock \bibinfo{title}{Plasmon-exciton-polariton lasing}.
\newblock \emph{\bibinfo{journal}{Optica}} \textbf{\bibinfo{volume}{4}},
  \bibinfo{pages}{31--37} (\bibinfo{year}{2017}).

\bibitem{Ha2018}
\bibinfo{author}{Ha, S.~T.} \emph{et~al.}
\newblock \bibinfo{title}{{Directional lasing in resonant semiconductor
  nanoantenna arrays}}.
\newblock \emph{\bibinfo{journal}{Nat. Nanotechnol.}}
  \textbf{\bibinfo{volume}{13}}, \bibinfo{pages}{1042--1047}
  (\bibinfo{year}{2018}).

\bibitem{Pourjamal2019}
\bibinfo{author}{Pourjamal, S.} \emph{et~al.}
\newblock \bibinfo{title}{{Lasing in Ni nanodisk arrays}}.
\newblock \emph{\bibinfo{journal}{ACS Nano}} \textbf{\bibinfo{volume}{13}},
  \bibinfo{pages}{5686--5692} (\bibinfo{year}{2019}).

\bibitem{Maccaferri2015}
\bibinfo{author}{Maccaferri, N.} \emph{et~al.}
\newblock \bibinfo{title}{{Ultrasensitive and label-free molecular-level
  detection enabled by light phase control in magnetoplasmonic nanoantennas}}.
\newblock \emph{\bibinfo{journal}{Nat. Commun.}} \textbf{\bibinfo{volume}{6}},
  \bibinfo{pages}{6150} (\bibinfo{year}{2015}).

\bibitem{FreireFernandez2020}
\bibinfo{author}{Freire-Fern{\'{a}}ndez, F.}, \bibinfo{author}{Mansell, R.} \&
  \bibinfo{author}{{van Dijken}, S.}
\newblock \bibinfo{title}{{Magnetoplasmonic properties of perpendicularly
  magnetized [Co/Pt]$_N$ nanodots}}.
\newblock \emph{\bibinfo{journal}{Phys. Rev. B}}
  \textbf{\bibinfo{volume}{101}}, \bibinfo{pages}{054416}
  (\bibinfo{year}{2020}).

\bibitem{Zou2004a}
\bibinfo{author}{Zou, S.}, \bibinfo{author}{Janel, N.} \&
  \bibinfo{author}{Schatz, G.~C.}
\newblock \bibinfo{title}{{Silver nanoparticle array structures that produce
  remarkably narrow plasmon lineshapes}}.
\newblock \emph{\bibinfo{journal}{J. Chem. Phys.}}
  \textbf{\bibinfo{volume}{120}}, \bibinfo{pages}{10871--10875}
  (\bibinfo{year}{2004}).

\bibitem{Guan2020}
\bibinfo{author}{Guan, J.} \emph{et~al.}
\newblock \bibinfo{title}{{Quantum dot-plasmon lasing with controlled
  polarization patterns}}.
\newblock \emph{\bibinfo{journal}{ACS Nano}} \textbf{\bibinfo{volume}{14}},
  \bibinfo{pages}{3426--3433} (\bibinfo{year}{2020}).

\bibitem{Lawley1980}
\bibinfo{author}{Lawley, K.~P.}
\newblock \emph{\bibinfo{title}{{Advances in Chemical Physics}}},
  vol.~\bibinfo{volume}{42} (\bibinfo{publisher}{John Wiley {\&} Sons, Inc.},
  \bibinfo{address}{Hoboken, NJ, USA}, \bibinfo{year}{1980}).

\bibitem{Pineider2013}
\bibinfo{author}{Pineider, F.} \emph{et~al.}
\newblock \bibinfo{title}{{Circular magnetoplasmonic modes in gold
  nanoparticles}}.
\newblock \emph{\bibinfo{journal}{Nano Lett.}} \textbf{\bibinfo{volume}{13}},
  \bibinfo{pages}{4785--4789} (\bibinfo{year}{2013}).

\bibitem{Kataja2016b}
\bibinfo{author}{Kataja, M.}, \bibinfo{author}{Pourjamal, S.} \&
  \bibinfo{author}{van Dijken, S.}
\newblock \bibinfo{title}{{Magnetic circular dichroism of non-local surface
  lattice resonances in magnetic nanoparticle arrays}}.
\newblock \emph{\bibinfo{journal}{Optics Express}}
  \textbf{\bibinfo{volume}{24}}, \bibinfo{pages}{3562} (\bibinfo{year}{2016}).

\bibitem{Vedmedenko_2020}
\bibinfo{author}{Vedmedenko, E.~Y.} \emph{et~al.}
\newblock \bibinfo{title}{The 2020 magnetism roadmap}.
\newblock \emph{\bibinfo{journal}{J. Phys. D: Appl. Phys.}}
  \textbf{\bibinfo{volume}{53}}, \bibinfo{pages}{453001}
  (\bibinfo{year}{2020}).

\bibitem{Lambert2014}
\bibinfo{author}{Lambert, C.-H.} \emph{et~al.}
\newblock \bibinfo{title}{All-optical control of ferromagnetic thin films and
  nanostructures}.
\newblock \emph{\bibinfo{journal}{Science}} \textbf{\bibinfo{volume}{345}},
  \bibinfo{pages}{1337--1340} (\bibinfo{year}{2014}).

\bibitem{Ota2020}
\bibinfo{author}{Ota, Y.} \emph{et~al.}
\newblock \bibinfo{title}{{Active topological photonics}}.
\newblock \emph{\bibinfo{journal}{Nanophotonics}} \textbf{\bibinfo{volume}{9}},
  \bibinfo{pages}{547–567} (\bibinfo{year}{2020}).

\bibitem{Wang2020c}
\bibinfo{author}{Wang, D.} \emph{et~al.}
\newblock \bibinfo{title}{{Lasing from finite plasmonic nanoparticle
  lattices}}.
\newblock \emph{\bibinfo{journal}{ACS Photonics}} \textbf{\bibinfo{volume}{7}},
  \bibinfo{pages}{630--636} (\bibinfo{year}{2020}).

\bibitem{Kichin2012}
\bibinfo{author}{Kichin, G.} \emph{et~al.}
\newblock \bibinfo{title}{{Metal-dielectric photonic crystal superlattice: 1D
  and 2D models and empty lattice approximation}}.
\newblock \emph{\bibinfo{journal}{Physica B Condens. Matter}}
  \bibinfo{pages}{4037--4042} (\bibinfo{year}{2012}).

\bibitem{Guo2017}
\bibinfo{author}{Guo, R.}, \bibinfo{author}{Hakala, T.~K.} \&
  \bibinfo{author}{T{\"{o}}rm{\"{a}}, P.}
\newblock \bibinfo{title}{{Geometry dependence of surface lattice resonances in
  plasmonic nanoparticle arrays}}.
\newblock \emph{\bibinfo{journal}{Phys. Rev. B}} \textbf{\bibinfo{volume}{95}},
  \bibinfo{pages}{155423} (\bibinfo{year}{2017}).

\bibitem{Cherqui2019}
\bibinfo{author}{Cherqui, C.}, \bibinfo{author}{Bourgeois, M.~R.},
  \bibinfo{author}{Wang, D.} \& \bibinfo{author}{Schatz, G.~C.}
\newblock \bibinfo{title}{{Plasmonic surface lattice resonances: Theory and
  computation}}.
\newblock \emph{\bibinfo{journal}{Acc. Chem. Res.}}
  \textbf{\bibinfo{volume}{52}}, \bibinfo{pages}{2548--2558}
  (\bibinfo{year}{2019}).

\bibitem{daskalakis_ultrafast_2018}
\bibinfo{author}{Daskalakis, K.~S.}, \bibinfo{author}{V{\"a}kev{\"a}inen,
  A.~I.}, \bibinfo{author}{Martikainen, J.-P.}, \bibinfo{author}{Hakala, T.~K.}
  \& \bibinfo{author}{T\"orm\"a, P.}
\newblock \bibinfo{title}{Ultrafast {{Pulse Generation}} in an {{Organic
  Nanoparticle}}-{{Array Laser}}}.
\newblock \emph{\bibinfo{journal}{Nano Lett.}} \textbf{\bibinfo{volume}{18}},
  \bibinfo{pages}{2658--2665} (\bibinfo{year}{2018}).

\bibitem{des2016plasmonic}
\bibinfo{author}{des Francs, G.~C.} \emph{et~al.}
\newblock \bibinfo{title}{Plasmonic {P}urcell factor and coupling efficiency to
  surface plasmons. {I}mplications for addressing and controlling optical
  nanosources}.
\newblock \emph{\bibinfo{journal}{J. Opt.}} \textbf{\bibinfo{volume}{18}},
  \bibinfo{pages}{094005} (\bibinfo{year}{2016}).

\bibitem{Zvezdin1997}
\bibinfo{author}{Zvezdin, A.~K.} \& \bibinfo{author}{Kotov, V.~A.}
\newblock \emph{\bibinfo{title}{{Modern Magnetooptics and Magnetooptical
  Materials}}} (\bibinfo{publisher}{Taylor and Francis Group},
  \bibinfo{address}{New York, NY}, \bibinfo{year}{1997}).

\bibitem{Sato_1992}
\bibinfo{author}{Sato, K.} \emph{et~al.}
\newblock \bibinfo{title}{{Magnetooptical spectra in Pt/Co and Pt/Fe
  multilayers}}.
\newblock \emph{\bibinfo{journal}{Jap. J. Appl. Phys.}}
  \textbf{\bibinfo{volume}{31}}, \bibinfo{pages}{3603--3607}
  (\bibinfo{year}{1992}).

\bibitem{JohnsonChristy}
\bibinfo{author}{Johnson, P.~B.} \& \bibinfo{author}{Christy, R.~W.}
\newblock \bibinfo{title}{Optical constants of the noble metals}.
\newblock \emph{\bibinfo{journal}{Phys. Rev. B}} \textbf{\bibinfo{volume}{6}},
  \bibinfo{pages}{4370} (\bibinfo{year}{1972}).

\end{thebibliography}

\section*{Acknowledgements}
This work was supported by the Academy of Finland (Grant Nos. 303351, 307419, 316857, 327293) and by the Centre for Quantum Engineering (CQE) at Aalto University. F.F-F. acknowledges financial support from the Finnish Academy of Science and Letters (Vilho, Yrjö and Kalle Väisälä Fund). Lithography was performed at the OtaNano - Micronova Nanofabrication Centre, supported by Aalto University. K.S.D acknowledges financial support from the European Research Council (ERC) under the European Union’s Horizon 2020 research and innovation programme (Grant No. 948260). J.C. acknowledges support by the Academy of Finland under project No. 325608 (SPATUNANO). We acknowledge the computational resources provided by the Aalto Science-IT project.

\section*{Author contributions}

F.F.-F. and S.P. fabricated the samples and characterized the optical and magneto-optical response of the square and rectangular plasmonic lattices. F.F.-F. and S.P. performed the lasing experiments and K.S.D. oversaw these measurements. P.T., J.C., F.F.-F., J.-P.M., and K.A. worked on the theory analysis. P.T. and S.v.D. supervised the work. F.F.-F., P.T., and S.v.D. wrote the manuscript with input from all authors.\\

\section*{Competing interests}
The authors declare no competing interests.

\end{document}